# SKA as a powerful hunter of jetted Tidal Disruption Events

**I. Donnarumma**[*,1], **E. M. Rossi**[2], **R. Fender**[3], **S. Komossa**[4], **Z. Paragi**[5], **S. Van Velzen**[6], **I. Prandoni**[7]

[1] *INAF-IAPS, via Fosso del Cavaliere 100, 00133, Rome, Italy;* [2] *Leiden Observatory, Leiden University;* [3] *Physics and Astronomy, University of Southampton, Southampton SO17 1BJ, UK.;* [4]*Max-Planck-Institut fuür Radioastronomie, Auf dem Hügel 69, 53121 Bonn, Germany;* [5] *Joint Institute for VLBI in Europe, Postbus 2, NL-7990 AA Dwingeloo, the Netherlands;* [6] *IMAPP, Radbound University, PO Box 9010, 6500 GL Nijmegen, The Netherlands;* [7] *INAF-IRA Bologna, Via P. Gobetti 101, 40129 Bologna, Italy;*
*E-mail:* `immacolata.donnarumma@iaps.inaf.it`

Observational consequences of the tidal disruption of stars (TDEs) by supermassive black holes (SMBHs) can enable us to discover quiescent SMBHs and constrain their mass function. Moreover, observing jetted TDEs (from previously non-active galaxies) provides us with a new means of studying the early phases of jet formation and evolution in an otherwise "pristine" environment. Although several (tens) TDEs have been discovered since 1999, only two jetted TDEs have been recently discovered in hard X-rays, and only one, Swift J1644+57, has a precise localization which further supports the TDE interpretation. These events alone are not sufficient to address those science issues, which require a substantial increase of the current sample. Despite the way they were discovered, the highest discovery potential for *jetted* TDEs is not held by current and up-coming X-ray instruments, which will yield only a few to a few tens events per year. In fact, the best strategy is to use the Square Kilometer Array to detect TDEs and trigger multi-wavelength follow-ups, yielding hundreds candidates per year, up to $z \sim 2$. Radio and X-ray synergy, however, can in principle constrain important quantities such as the absolute rate of jetted TDEs, their jet power, bulk Lorentz factor, the black hole mass function, and perhaps discover massive black holes (MBH) with $< 10^5 M_\odot$. Finally, when comparing SKA results with information from optical surveys like LSST, one can more directly constrain the efficiency of jet production.

[*]Speaker.

# 1. Introduction

Mounting observational evidence is supporting a scenario where most galactic nuclei host supermassive black holes (SMBHs). Gas inflow from larger scales causes a small ($\sim 1\%$) fraction of SMBHs to accrete continuously for millions of years and shine as quasars. Most of them are instead "quiescent", accreting –if at all– at a highly sub-Eddington rate. Observationally, it is therefore hard to assess the presence and measure the mass of most of SMBHs, beyond the local universe. Occasionally, however, a sudden increase of the accretion rate may occur, if a large mass of gas, e.g. a star, falls into the tidal sphere of influence of the black hole and find its-self torn apart and accreted. We call these events "tidal disruption events" (TDEs). TDEs can result in sudden flares: these can reach the luminosity of a quasar but they are rare ($\sim 10^{-5} \text{yr}^{-1}$ per galaxy) and last for only a few up to several months. We will show in this Chapter how the SKA sensitivity and its large field of view could increase the TDE sample and have a large discovery potential.

The detection and study of these flares can deliver other important astrophysical information, beyond flagging the presence of a SMBH. The theoretical expectation is that TDE flares would be caused by sudden accretion of the star debris. If the star is completely disrupted, its debris is accreted at a decreasing rate of $\dot{M} \propto t^{-5/3}$ (Rees 1988; Phinney 1989). Therefore, TDEs allow us to study the formation of a transient accretion disc and its continuous transition through different accretion states. As the accretion rate decreases, we can in principle observe a disc in an initial super-Eddington phase, lasting several months; the disc then would become *slim* and later enter in a *thin* disc regime, finally ending its life, years later, in a radiative *inefficient* state. The super-Eddington phase — which occurs only for SMBH masses $M \lesssim 10^7 \text{ M}_\odot$ — is theoretically uncertain, but it may be associated with a powerful radiative driven wind (Rossi & Begelman 2009), which thermally emits $\sim 10^{41} - 10^{43}$ erg s$^{-1}$, mainly at optical frequencies (Strubbe & Quataert 2009; Lodato & Rossi 2011). The disc luminosity ($\sim 10^{44} - 10^{46}$ erg s$^{-1}$), instead, peaks in the far-UV/soft X-rays (Lodato & Rossi 2011). Of great theoretical importance would also be the possibility to observe the formation and evolution of an associated jet, powered by this sudden accretion episode.

The first TDEs were discovered in *ROSAT* surveys of the X-ray sky (e.g., Komossa & Bade 1999; Grupe et al. 1999) — see Komossa (2002) for a review. Later, *GALEX* allowed for the selection of TDE at UV frequencies (Gezari et al. 2009, 2012); many of the most recent TDE candidates are now found in optical transient surveys (van Velzen et al. 2011a; Cenko et al. 2012; Chornock et al. 2014; Arcavi et al. 2014). An alternative method to select TDE candidate is to look for optical spectra with extreme coronal lines (Komossa et al. 2008; Wang et al. 2012). These "thermal" features are believed to be associated with emission from the disc or the radiative driven wind.

Moreover, recently the *Swift* Burst Alert Telescope (BAT) instrument (Burrows et al. 2011; Cenko et al. 2012), triggered two TDE candidates in the hard X-ray band. A multi-frequency follow-up from radio to $\gamma$-rays revealed a new class of non-thermal TDEs. It is widely believed that emission from a relativistic jet is responsible for the hard X-ray spectrum (with power-law photon index $\beta \sim 1.6 - 1.8$) and the increasing radio activity (Levan et al. 2011), detected a few days after the trigger. The best studied of the two events is Swift J1644+57 (Sw J1644 in short), whose radio lightcurve (see next section) is used in this Chapter to estimate detection rates in the radio band. The two main features that support the claim that Sw J1644 is a tidal disruption event are i) the X-ray lightcurve behaviour, that follows $\propto t^{-5/3}$ after a few days from the trigger, and ii) the radio localization of the event within 150 pc from the centre of a known quiescent galactic nucleus (Zauderer et al. 2011).

### 1.1 Radio emission of jetted TDEs

In 2011, BAT discovered a peculiar gamma-ray burst, now known as Sw J1644. What first caught the attention was the flaring activity, which caused multiple BAT triggers, followed by an X-ray light curve that extended to much longer timescales than any known GRB (Levan et al. 2011). The source was confirmed to be of extragalactic origin, its peak luminosity $\sim 100$ times higher than bright active galactic nuclei (Burrows et al. 2011; Levan et al. 2011). Follow-up radio observations revealed a transient radio counterpart (Zauderer et al. 2011) and the multi-wavelength SED of the source resembled that of BL Lac objects (Bloom et al. 2011). Key evidence that Sw J1644 is, in fact, a newly launched jet, is the rapid X-ray variability implying Doppler-beamed emitting regions (Bloom et al. 2011). A second BAT-triggered TDE candidate, Swift 2058+05, that also has a transient radio counterpart was discovered by Cenko et al. (2012).

A rich dataset of radio follow-up observations of Sw J1644 is presented in Berger et al. (2012); Zauderer et al. (2013). At 5 GHz, the peak in the light curve occurs about 100 days after the first gamma-ray trigger, at $F_\nu = 15$ mJy ($\nu L_\nu = 4 \times 10^{41}$ erg s) (cf. Fig. 3 in Zauderer et al. 2013). During the first years the self-absorption frequency is around 10 GHz; at 2 GHz the peak of the light curve is expected to occurs about 500 days after the stellar disruption. The lightcurve at 1.4 GHz is shown in Fig. 3.2. Data were collected from 5 days after the Swift trigger, with a flux of $F_\nu \geq 0.2$ mJy. The radio light curve of the second relativistic TDE, Swift 2058+05, has been sampled for only a few epochs (at 5 GHz the flux is 0.9 mJy). Its luminosity is similar to Sw J1644, but the spectrum is rather flat ($F_\nu \propto \nu^0$) from 4.5 GHz up to 22.5 GHz Cenko et al. (2012).

The light curve of Sw J1644 can be modeled as synchrotron emission from the forward shock (Giannios & Metzger 2011; Metzger, Giannios & Mimica 2012), although this model requires an increase of the isotropic-equivalent jet kinetic energy (Berger et al. 2012), which is much stronger than what is expected from the fallback rate of stellar debris (see also Kumar et al. 2013). Interestingly, the X-ray light curve Sw J1644, shows a very steep drop when the (theoretically estimated) fallback rate crosses the Eddington limit (Zauderer et al. 2013). This could imply that the jet engine has switched off due to a state change of the accretion disk, as suggested by the first models of TDE jets (Giannios & Metzger 2011; van Velzen, Körding & Falcke 2011).

### 2. Constraints on TDE jets from past and present radio surveys

So far, surveys of radio transients have yielded a few events (e.g, Gal-Yam et al. 2006; Bannister et al. 2011) — see Frail et al. (2012) for a review. However, due to the lack of follow-up observations it has been difficult to test if these are TDE jets. For a given light curve model of the radio emission from TDE jets, the upper limits on the event rate from these surveys can be used to constrain the rate of these events.

As shown in van Velzen et al. (2013), the observed light curve of Sw J1644 can be used to make an order of magnitude estimate of snapshot rate of events that have a similar light curve.

Input to this estimate are the per-galaxy rate of tidal disruption jets ($\dot{N}_{\rm TDE}$) and the mean volume density of black holes $\rho_{\rm BH}$ that create observable TDE jets. If we assume that only jets with an inclination smaller than $1/\Gamma_j$ are visible, we arrive at the following expression for the snapshot rate above a given flux limit $F_{\nu,\rm lim}$:

$$R(F_{\nu,\rm lim}) \sim 8 \times 10^{-3} \, \Gamma_j^{-2} \left(\frac{F_{\nu,\rm Sw}}{F_{\nu,\rm lim}}\right)^{3/2} \frac{\Delta T \dot{N}_{\rm TDJ}}{10^{-5}} \frac{\rho_{\rm BH}}{5 \times 10^{-3} \, {\rm Mpc}^{-3}} \, {\rm deg}^{-2} \quad . \quad (2.1)$$

Here $\Delta T$ denotes the time interval along which Sw J1644 flux is above $F_{\nu,\rm lim}$. Here we normalized snapshot rate assuming that the rate of jetted TDE $\dot{N}_{\rm TDJ}$ is similar to the rate of thermal TDE (we return to this assumption below). The rate of thermal TDE can be estimated from soft X-ray (Donley et al. 2002) or optical (van Velzen & Farrar 2014) surveys, a conservative value is $\dot{N}_{\rm TDE} \sim 10^{-5} \, {\rm yr}^{-1}$.

For $\Delta T = 1$ yr, the light curve of Sw J1644 implies $F_{\nu,\rm Sw} = 10$ mJy, 2 mJy at 5 and 1 GHz, respectively. Adopting $\Gamma_j = 2$ as estimated from modelling the late-time radio emission of Sw J1644 (Berger et al. 2012), we obtain our estimate of snapshot rate as shown in Fig. 1. This rate is close to the existing upper limits of radio transient surveys.

In Fig. 1 we also show the predicted rate at 1 GHz based on the model of van Velzen, Körding & Falcke (2011). In the optimistic scenario of these authors, 20% of the accretion power is converted into jet power (i.e., in analogy with AGNs, tidal disruption jets are always radio-loud). The synchrotron luminosity is computed using the conical equipartition jet model (Blandford & Königl 1979; Falcke & Biermann 1995), which yields a peak luminosity (for on-axis events) that is lower than the peak of Sw J1644 (hence the predicted snapshot rate is also lower). In the conservative scenario of van Velzen, Körding & Falcke (2011), tidal disruption jets are only radio-loud when the fallback rate drops below 2% of the Eddington limit.

## 3. SKA as the best hunter of *jetted* TDEs

In this section we report on the rate of TDEs expected to be detected with SKA operating in survey mode. The SKA1-survey (SKA1-SUR) provides us with an optimal tool to catch and discover jetted TDEs. We adopted a strategy which relies on a trade-off between sky coverage and sensitivity, fully compliant with the SKA1-SUR science requirements reported in the Baseline Design Document (Dewdney et al. 2013). We request to achieve an half sky coverage (20,000 deg$^2$) with a 2-day cadence at a $5-\sigma$ flux limit of 90 $\mu$Jy. This can be achieved by SKA1-SUR. SKA1-SUR can cover this area with 1111 pointings at 1.4 GHz (the SKA1-SUR field of view at this frequency is 18deg$^2$) at a sensitivity of 3.72 microJy hr$^{-1/2}$ (see Table 1 in Dewdney et al. 2013), or at a flux limit of 90 $\mu$Jy in 2 days.

### 3.1 Snapshot rate estimate

We first make a rough estimate based on Eq. 2.1. At 1 GHz, the observed light curve of Sw J1644 is above 2 mJy for less than one year ($\Delta T \approx 0.8$yr) (Zauderer et al. 2013), for a flux limit 90 $\mu$Jy we thus obtain a snapshot rate of 0.2 deg$^{-2}$ (Fig. 1), implying $\approx 4000$ events for the SKA transient survey adopted here if $\Gamma_j = 2$ is assumed.

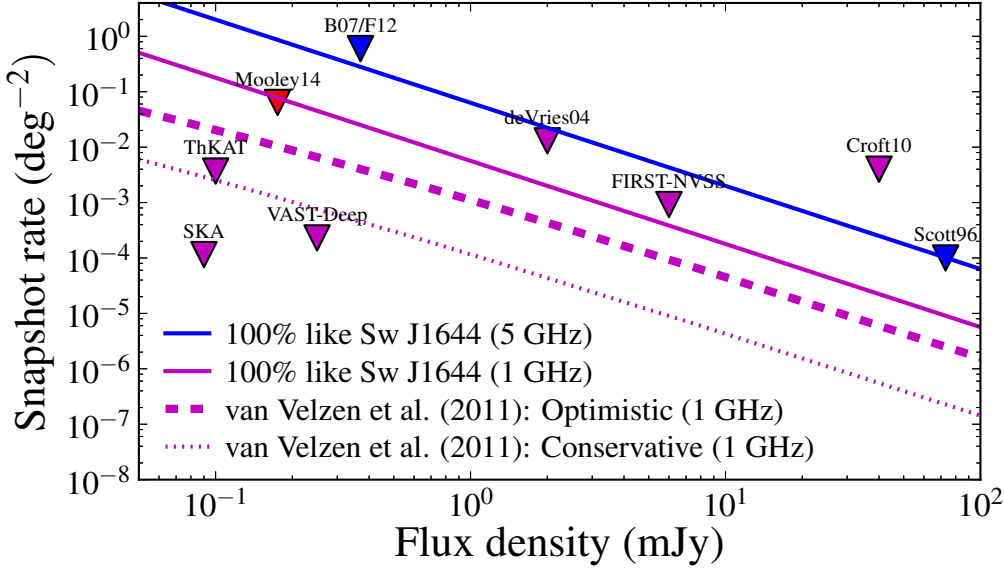

**Figure 1:** Predicted snapshot rate and upper limits from current and future radio surveys. The predicted rate for Sw J1644 is based on eq. 2.1 and assumes a 100% jet production efficiency, i.e., every TDE makes a jet with a light curve similar to Sw J1644. The blue, red, and magenta triangles show (potential) 2-$\sigma$ upper limits at 5, 3 and 1 GHz, respectively. References: B7/F12: Bower et al. (2007); Frail et al. (2012); de Vries et al. (2004), FIRST-NVSS: Gal-Yam et al. (2006); Croft et al. (2010); Scott (1996); Mooley et al. (2014).

### 3.2 Radio Lightcurve and Monte Carlo simulations

We here extend our rate calculations to explicitly account for redshift, BH mass M and stellar mass $m_*$ dependences, using Monte Carlo simulations (see Donnarumma & Rossi 2014, for details). The radio lightcurve at 1.4 GHz of Sw J1644 is plotted in Fig. 3.2. To investigate the detection capability of SKA1-SUR, we assume that all events have the same light curve shape as Sw J1644. We model the mission as synchrotron radiation and use the peak flux and characteristic frequencies measured at different times by Berger et al. (2012, see their eq.4 and dash-dotted line in our Fig.3.2) extended at late epochs by Zauderer et al. (2013). The lightcurve behaviour in the first $\approx 8$ days (where there are no data) has been extrapolated from later epochs. The result is a quite sharp rise, that can be considered a conservative assumption: a more gentle rise would ensure more detections. Note that ours is *not a fit* to the data, but *it is a modelling*. From Fig. 2 it is clear that the radio flux does not show the temporal characteristic behaviour $\propto t^{-5/3}$ of a tidal disruption event, that it is present in the X-ray band. Therefore, a follow up strategy at higher frequency is required for identification (see Section 3.5).

An appropriate spectral correction can describe events which explode at different redshifts than Sw J1644 [1]. Rescaling the flux for different BH and stellar masses requires instead to assume a jet model. A first possibility is to describe the jet evolution with a Blandford McKee solution (Blandford & Mckee 1977), usually adopted for $\gamma$-ray burst afterglows. In this case, the radio flux does *not* explicitly depend on the BH mass. A second approach is to assume that the luminosity is proportional to the peak of the jet power, which in turn is proportional to the peak fall-back rate $L_{j,p} \propto \dot{M}_p$. This introduces a mass dependence in the luminosity as $L_\nu \propto L_{j,p} \propto M^{-1/2} m_*^{1/2}$. Hereafter we will focus on this second BH mass dependent lumimosity (MDL) model and we refer the readers to Donnarumma & Rossi (2014) for a comparison with the Blandford Mckee solution. The stellar mass distribution in galactic nuclei may vary, but there is no solid observational evidence. We therefore assume a "universal" Kroupa initial mass function Kroupa (2001), as observed in our Galaxy. The BH mass function is instead explained more in detail in the following section.

---

[1] We here assume that all jets have approximately the same Doppler factor $\delta \approx \Gamma_j \simeq 2$, as measured in radio for Sw J1644

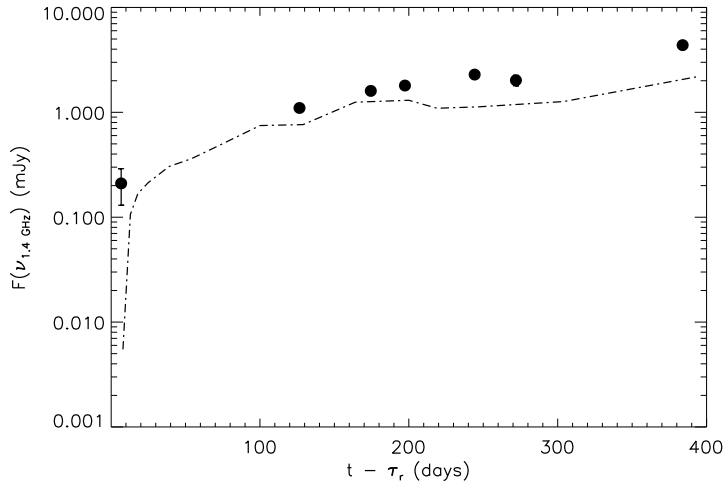

**Figure 2:** The radio (1.4 GHz) light curve of Sw J1644: data (black circles) from Berger et al. (2012); Zauderer et al. (2013), modelling (dash-dotted black line) from Donnarumma & Rossi (2014). Figure adapted from Donnarumma & Rossi (2014). Note that the black line is not a fit. We trace it by assuming synchrotron emission and the characteristic frequencies and peak flux measured by Berger et al. (2012); Zauderer et al. (2013) at different times. The result slightly underpredicts the data, but recovers the overall temporal behaviour which is rather flat.

### 3.3 Black hole mass functions

The mass distribution of black holes as a function of redshift is a key ingredient to calculate TDE rates. We adopt the BH mass function (MF) predictions according to the models labeled *G* and *G(z)* in Shankar, Weinberg & Miralda-Escudé (2013). In practice, we consider the two accretion models which yield the largest and the lowest MF predictions in order to reflect the uncertainty due to the black hole mass distribution on our expected TDE rates. The "intrinsic" TDE rate as a function of redshift is defined by

$$R(z) = \int_{M_{\min}}^{M_{\max}} \phi(M,z) \dot{N}_{\mathrm{TDE}} V(z) dM, \qquad (3.1)$$

where $\dot{N}_{\mathrm{TDE}} = 10^{-5}$ yr$^{-1}$ is as our fiducial *intrinsic* TDE rate, $\phi(M,z)$ is the black hole comoving number density and $V(z)$ the comoving cosmological volume. In our calculations, we consider only BH masses above $M_{\rm min} = 10^6$ M$_\odot$, because MFs are not observationally constrained below that mass.

### 3.4 Monte Carlo simulations

Assuming the radio modellings described in section 3.2, we performed Monte Carlo simulations (MCs) to derive the rate of jetted TDEs expected to be detected by SKA1-SUR operating in wide survey mode .

The intrinsic rate $R(z)$ (eq. 3.1) is reduced by a factor of $1/2\,\Gamma_j^2$ for the relativistic beaming. Our fiducial value is $\Gamma_j = 2$ (Zauderer et al. 2013; Berger et al. 2012), but our results may be easily rescaled by assuming different values of $\Gamma_j$. Finally, a further scaling factor accounts for the fraction of the sky accessible at each time by the instrument. As already discussed in section 3.3, we will provide a range of rates whose limits reflect the uncertainty in the BH MF predictions.

The specific observational strategy for SKA1-SUR explained at the beginning of this section assumes an half sky coverage (20,000 deg$^2$) with a 2-day cadence at a $5-\sigma$ flux limit of 90 $\mu$Jy.

For each event, with given redshift, BH and stellar mass, the light curve is predicted as described in section 3.1. We extract the trigger time randomly from a 1 year uniform distribution, and we calculate the average flux over 2 days since the trigger and then we compare it with the SKA1-SUR flux limit fixed for our strategy.

We show in Fig. 3, the distribution of the TDE rate as a function of $z$ (right panel) and the corresponding BH mass distribution (left panel) for the two BH MFs assumed (solid and dotted blue lines). In the right panel, we overplot the rate distributions for TDEs with BH masses lower than $10^7$ M$_\odot$ (solid and dotted red lines). The total rate distribution peaks around $z \sim 0.5$ and it is dominated by light SMBHs locally and for $z$ above the peak. The maximum redshift that SKA can attain is close to $z \sim 2$.

The total number of objects obtained by integrating these distributions in $z$ and M$_{BH}$ is $\sim 300$ and $\sim 800$ yr$^{-1}$ for the G and G(z) models, respectively. We note that our integrated rates are significantly lower than the first simple estimates reported in section 3. This is mainly due to the fact that we accounted for the stellar mass dependence, that introduces a large variation in the intrinsic luminosity of these events. Nevertheless, our results imply that a significant increase in the number of detected jetted TDEs can be achieved thanks to our strategy in the framework of SKA1-SUR. However, as mentioned before, the TDE signature seems to appear at higher frequencies. We therefore proceed to consider X-ray facilities.

### 3.5 Radio and X-ray synergy

Future X-ray surveys can detect much fewer candidates than SKA. The e-Rosita all sky survey (Merloni et al. 2012) is expected to trigger a few jetted TDEs by using its "hard" X-ray band ($2-10$ keV). Calculations based on the Swift statistics (by using SW 2058 as a prototype) predicted 1 object to be detected per scan (6-month long) probing the universe up to $z \sim 4.5$ (Khabibullin et al. 2014).

We also envisage the use of X-ray telescopes to follow up radio candidates and confirm their nature. To quantitatively explore this possibility, we model the X-ray emission of Sw J1644 in the 1-10 keV band, assuming that the X-ray luminosity scales as the fall back rate, $L_X \propto \dot{M}$. (see

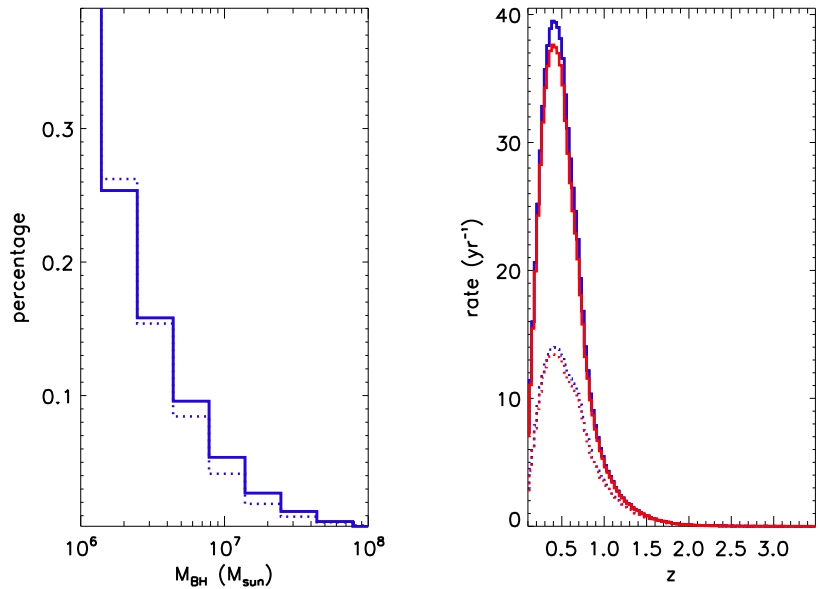

**Figure 3:** Rate of events detected by SKA as a function of redshift (right panel) and their mass distribution (expressed in percentage, left panel), for two different black hole mass distribution functions (solid and dotted blue lines). The rates for BH masses lower than $10^7$ M$_\odot$ are also shown (red lines, right panel). Figure adapted from Donnarumma & Rossi (2014).

Donnarumma & Rossi 2014, for a full description of the methodology). In our estimates, we assumed that a follow-up can be arranged within 1 day from the radio trigger. To claim detection, we require that the observed flux can be detected with a signal to noise of at least 10 during four pointings, spread over four days. This requirement implies that a flux threshold (unabsorbed flux) of $\sim 5 \times 10^{-12}$ erg cm$^{-2}$ s$^{-1}$ in the 1-10 keV band is in principle enough to identify all the SKA selected candidates.

Therefore, given their sensitivity, future X-ray experiments like Athena (Nandra et al. 2013) and a LOFT-like mission (Feroci et al. 2012) could offer a unique chance to follow-up and characterize SKA-triggered TDEs. If Athena will have a follow-up efficiency of 50% associated with the sky accessibility, it will be able to identify about 50% of all the radio sample. For a LOFT-like mission, we would expect that a sub-sample of radio TDEs between $\simeq 100$ and $\simeq 350$ will be detected per year up to $z \sim 1.2$ by observing with its main instrument, the Large Area Detector.

## 4. Constraints on the jet production efficiency

As recalled above, only two jetted TDEs have been detected, while detections of the thermal candidates, related to the presence of an accretion disk, have been more numerous. The question then arises whether this is due to observational biases or to a small jet opening angle or to an intrinsic low efficiency of transient accretion disks to produce (luminous) jets.

To answer this question, one may first compare predictions in the Swift/BAT bands with the observed rate of $\approx 0.3$ yr$^{-1}$. It turns out that the most pessimistic predictions by Donnarumma & Rossi (2014) are still a factor of tens higher than observed and it would be tempting to reconcile this discrepancy by invoking a jet production efficiency of a few percent. However, it has been shown that this inference *cannot* be confidently drawn (Donnarumma & Rossi 2014). On the one hand, other uncertain parameters –like the value of the beaming factor in the X-ray emitting region – can mitigate the discrepancy. On the other, the modelling of the characteristics (e.g. flux limit and sky coverage) of an effective Swift/BAT survey (in fact, Swift is *not* operating in survey mode) it is not at all straightforward and does not allow for firm predictions. Plus, with just two events, the Poisson uncertainty is very large.

Another approach for estimating the jet production efficiency is to use instead the observed thermal candidates. Thermal TDEs may also contain jets. In fact, based on an analogy with X-ray binaries, one may anticipate that each TDE is accompanied by a jet (see van Velzen, Körding & Falcke 2011, and references therein). Because they are observed off-axis, these jets are less luminous and will not trigger *Swift*, but they can be detected through radio or X-ray follow-up observations.

The existing radio follow-up observations of known thermal TDEs are summarized in van Velzen et al. (2013). So far, none of the optical/UV-selected TDE have been detected at radio frequencies (see also Arcavi et al. 2014). Using a simple model to scale the observed light curve of Sw J1644 for off-axis observed we find that the current radio observations are deep enough to detect jets similar to Sw J1644 to an inclination of $i < 70°$ (van Velzen et al. 2013). We thus know that the jet production efficiency has to be less than 100%.

Two of the seven TDE candidates that were detected by ROSAT or XMM have been detected at 5 GHz in follow-up observation by Bower et al. (2013), namely IC 3599 (Komossa & Bade 1999) and RX J1420+5334 (Greiner et al. 2000). The radio observations were obtained about ten years after the estimated time of disruption, their transient nature remains to be confirmed (in fact one of the radio-detected sources, IC 3599, is more likely to be an AGN).

We can thus conclude that none of the thermal TDE show strong evidence for relativistic jets. This could imply that the fraction of TDE that launch jets similar to Sw J1644 less than $\approx 10\%$, although currently is not clear if the detection of optical/UV emission implies a bias against the detection of radio emission. To be able to detect optical or UV emission, one requires a circum-nuclear environment with little absorption by dust, while the radio luminosity can be boosted by a high density circumnuclear gas.

Present constraints on the jet production efficiency suffer of the poor statistics due to the small X-ray and radio samples. The radio rates of jetted TDEs predicted in this work, will help to overcome the present limitation due to the very large sample expected to be triggered by the SKA. Besides serving as a trigger, the high survey speed of the SKA implies that it will be possible to have near-simultaneous observations with LSST surveys. The detection rate of optical TDE with LSST is $\sim 1000$ yr$^{-1}$ (e.g., Gezari et al. 2009; van Velzen et al. 2011a), hence the SKA observations will be able to measure a jet production efficiency for thermal TDE as small as 0.1%. Comparing the SKA-triggered TDEs with the SKA observations of thermal TDE discovered by LSST will provide a new window to study the jet physics of massive black holes.

## 5. Implications for the cosmological growth of SMBHs

The observed properties of central SMBH and their hosts show evidence for a strong link between SMBH and galaxy formation (Ferrarese & Merritt 2000; Gebhardt et al. 2000). Outstanding questions include how the massive black hole seeds (MBH, $\sim 10^2 - 10^5$ $M_\odot$) were initially formed, and their successive mass growth, which determine their mass functions (Volonteri 2010). TDEs will provide a great opportunity in this quest.

Fig.3 shows that SKA will more easily detect light ($< 10^7 M_\odot$) MBHs locally and at z > 1. Since the lighter the more luminous, this result suggests the possibility to observe BHs in the elusive "intermediate mass regime" in the distant Universe.

As described in section 3, in this Chapter we focused our work on BHs with masses $\geq 10^6 M_\odot$. This because of the very high uncertainties affecting the lower mass side of the BH mass functions. However, the recently reported TDE candidate from a $\sim 10^{5.5}$ $M_\odot$ black hole (Donato et al. 2014) shows the great potential to exploit TDEs to seek out MBHs in the mass range of $\sim 10^2 - 10^5$ $M_\odot$, with important consequences for BH formation models. These black holes must exist in low-mass dwarf galaxies or perhaps as remnants inside massive galaxies. Alternatively, they may have formed in the centre of globular clusters[2] (Maccarone 2004).

We would like to point out, that the high-precision location of the radio-TDE with respect to the galaxy core will be of great relevance in the understanding of both SMBH and IMBH formation. TDEs might reveal recoiling black holes (e.g. Komossa 2012) that have wandered slightly away from the nucleus, off-nuclear SMBH due to ongoing merger activity (Comerford & Greene 2014), or for example black holes that were formed in the centre of globular clusters (Maccarone 2004). In the case of galaxy mergers, the stellar tidal disruption rate can be temporarily strongly enhanced (e.g., Chen et al. 2009). Such TDE would point us to an ongoing merger, but would likely be off-set from the geometrical center of the merging system. In the case of recoiling BHs due to e.g. SMBH coalescence, gravitational wave recoil will insert a kick velocity on the newly formed single black hole, which will carry a retinue of stars with it, as it is ejected from the core of its host galaxy. These bound stars will undergo tidal disruption (Komossa & Merritt 2008), leading to a radio event offset from the galaxy core.

## 6. Observational strategies

As showed in Section 3, SKA can play an important role in discovering jetted TDEs. To achieve this, it will be necessary to find TDEs by means of a commensal search in all SKA observations, or by carrying out a near real-time transient search in a high cadence dedicated survey observations. The latter is preferred since our goal is to detect TDEs early on in order to allow for the necessary follow-ups (to confirm the TDE origin). The real advantage of this strategy is that it will be able to build up a large sample of jetted TDEs up to $z \sim 2$, where the low-mass population of SMBHs could dominate, providing a unique opportunity to constrain the low mass end of the SMBH mass function in the far Universe.

---

[2]Off-nuclear BH in the mass range $\sim 10^2 - 10^5$ $M_\odot$, as a likely power source of the most extreme ultra-luminous X-ray sources, and the ones that are believed to have formed in the centre of globular clusters are often referred to as intermediate-mass black holes (IMBH) in the literature.

The SKA1-SUR will provide us with an optimal instrument to detect transient events as TDEs. Our strategy, explained in Section 3, is compliant with the Baseline Design Document (Dewdney et al. 2013) and predicts a few hundred detections with X-ray follow up for identification. However, a significant sample could also be collected, assuming a 50% reduction in sensitivity due to re-baselining. This is because the 1.4 GHz light curve of Sw J1644 increases with time at least up to 600 days (Zauderer et al. 2013), therefore a multiple cadence survey with longer integration times ($\sim 8$ days) needed to achieve the same flux limit of our strategy should not result in a significant loss of radio detections. However, radio triggers at late times could be problematic for follow-up observations, because sources would get fainter in both optical and X-ray energy range, preventing the identification of higher redshift TDEs. Nevertheless, the bulk of the sample at $z \lesssim 0.5$ should still be successfully followed up at higher frequencies.

On the basis of these considerations, TDE can be commensally searched for, by exploiting the SKA1-SUR all-sky survey (31,000 deg$^2$ at 2 $\mu$Jy beam$^{-1}$ at 1.5-2 arcsec resolution) proposed in the framework of both the Magnetism and Continuum science cases (see Johnston-Hollit et al. 2015; Prandoni et al. 2015). If carried out by visiting individual fields multiple times with a cadence of $\sim 10$ days over the two year period, it will be very efficient in finding extra-galactic transients down to $\sim$90 $\mu$Jy beam$^{-1}$.

As mentioned before, the 1.4 GHz light curve of our prototype TDE does not bear any specific footprints of a TDE nature. Thus we cannot use only its properties or variability to distinguish a TDE candidate from neither a slowly variable AGN or a GRB. However, we can account for AGN contamination by cross-correlating the transient positions with deep AGN catalogues. A significant advance in this direction will be achieved by the large amount of data that LSST will provide in the near future together with those collected after one year of SKA and SKA precursor (ASKAP) observations. On the other hand, precise localization of the radio transient in the core of galactic nuclei will help assessing the nuclear origin: the better the localisation, the stronger is the case for a TDE.

This means that a rapid optical follow-up aimed at determining the host galaxy has to be foreseen. Once the host galaxy is found, thanks to a resolution of about 2 arcsec (SKA1-SUR) and 0.6 arcsec or better (SKA1-MID) (Dewdney et al. 2013) it will be possible to localize the brighter transients with a precision of $\sim 100$ milliarcsecond (mas), essential to separate nuclear transients from other phenomena (e.g., GRB). This angular localization is at the current alignment accuracy level of radio and optical reference frames (Orosz & Frey 2013), corresponding to linear scales of 0.5–0.8 kpc at $z > 0.3$.

We note that by detecting radio emission associated with star formation, pre-flare SKA observations can help to identify jetted TDEs. By combining the individual snapshot obtained *after* $\sim 1$ year of observations, we expect to be able to see typical star-forming galaxies ($L \sim 10^{20}$ W/Hz; Condon (1992)) up to $z \approx 0.5$. At this redshift we also expect to see most of our detected jetted TDEs (see Fig. 2), hence most TDEs that occur in star-forming galaxies will have a detected host galaxy. These host detections will allow for an accurate localization of the radio transient with respect to the center of its host galaxy, which will help to reject SNe and GRBs (TDE from SMBHs occur in the nucleus of the galaxy, while SNe/GBRs track the stellar density). The identification of nuclear radio transients will allow for faster triggers for follow-ups in X-rays (by avoiding the need for the optical follow-up for the host identification), which will give us a firm signature of the TDE nature if the trigger occur within 1 year from the beginning of the event (see section 3). A synergy with SKA and VLBI will help to characterize the jet formation in the pristine environment of the TDE hosts (Paragi et al. 2014b).

Given the huge amount of TDE candidates predicted and the *expensive* strategy, we do not need to know immediately for all hundreds of events, what they are. In the first stage, we could concentrate the efforts on the brightest events (tens of objects) and for them activate follow-up observations at higher energies. The rest of those events could be analyzed later by means of commensal search in SKA observations.

Finally, SKA will play an important role even for the follow-up of all TDEs detected by the *Gaia* mission (expected to end soon before SKA1-MID early operations) thanks to the SKA1-MID sub-arcsec angular resolution. Immediate follow-up observations of TDEs detected in the X-rays and by LSST in the optical will be important to constrain the jet production efficiency at low redshift. This will require flexible operations and the ability to respond to triggers from other instruments (see Fender 2015).

## 7. Conclusions

In this Chapter, we have shown how SKA alone, thanks to its combination of survey speed (SKA1-SUR, for transient searches) and sub-arcsec angular resolution (SKA1-MID, for TDE identifications), and in synergy with coeval instruments at higher energies, can play a leading role in discovering jetted TDEs and allowing for the first time statistical studies of their features. We predict a sample of hundreds, up to redshift $\sim 2$. It is however, important to be able to communicate the detection promptly, so that an immediate campaign to localize the event and trigger higher frequency instruments may be launched within a week from the TDE discovery. A lot of relevant information will be gathered from TDE discoveries thanks to the synergy among future radio, optical and X-ray surveys, allowing to:

- make the estimate of the jet production efficiency more reliable

- differentiate different theories of jets production onto supermassive black holes, e.g., spin-powered versus accretion-disk powered

- discover quiescent SMBHs in the distant Universe

- probe the BH mass function at low masses.

## References

Arcavi, I., Gal-Yam, A., Sullivan, M., et al. 2014, ArXiv e-prints, arXiv:1405.1415

Bannister, K. W., Murphy, T., Gaensler, B. M., Hunstead, R. W., & Chatterjee, S. 2011, Mon. Not. Roy. Astro. Soc., 412, 634

Barth, A. J., Ho, L. C., Rutledge, R. E. & Sargent, W. L. W. 2004, Ap. J., 607, 90

Berger, E., Zauderer, A., Pooley, G. G., et al. 2012, Ap. J., 748, 36

Blandford, R. D., & Königl, A. 1979, Ap. J., 232, 34

Blandford, R.. D., & McKee, C. F., 1977, Mon. Not. Roy. Astro. Soc., 180, 343
Bloom, J. S., Giannios, D., Metzger, B. D., et al. 2011, Sci, 333, 203
Bower, G. C. 2011, Ap. J., 732, L12
Bower, G. C., Metzger, B. D., Cenko, S. B., et al. 2013, Ap. J., 763, 84
Bower, G. C., Saul, D., Bloom, J. S., et al. 2007, Ap. J., 666, 346
Burrows, D. N., Kennea, J. A., Ghisellini, G., et al. 2011, Nat, 476, 421
Campana, S., Lodato, G., D'Avanzo, P., et al. 2011, Nat, 480, 69
Cannizzo, J. K., Troja, E., & Lodato, G. 2011, Ap. J., 742, 32
Carilli, C. L., Rawlings, S. 2004, New Astronomy Reviews, 48, 979
Cenko, S. B., Krimm, H. A., Horesh, A., et al. 2012, Ap. J., 753, 77
Chen, X., Madau, P., Sesana, A., Liu, F. K.. 2009, Ap. J., 697, 149
Chornock, R., Berger, E., Gezari, S., et al. 2014, Ap. J., 780, 44
Comerford, J. M., & Greene, J. E. 2014, Ap. J., 789, 112C
Condon, J. J. 1992, ARA& A, 30, 575
Croft, S., Bower, G. C., Ackermann, R., et al. 2010, Ap. J., 719, 45
de Vries, W. H., Becker, R. H., White, R. L., & Helfand, D. J. 2004, Astron. J., 127, 2565
Dewdney, P., Turner, W., Millenaar, R., McCool, R., Lazio, J., Cornwell, T., 2013, "SKA1 System Baseline Design", Document number SKA-TEL-SKO-DD-001 Revision 1
Donato, D., Cenko, S. B., Covino, S., et al. 2014, Ap. J., 781, 59
Donley, J. L. ,Brandt, W. N., Eracleous, M., & Boller, T. 2002, Ap. J., 124, 1308
Donnarumma, I., & Rossi, E. M. 2014, Ap. J., submitted
Fender, R. P., et al. 2015, "The Transient Universe with the Square Kilometre Array", in proceedings of "Advancing Astrophysics with the Square Kilometre Array", PoS(AASKA14)051
Feroci, M., Stella, L., van der Klis, M., et al. 2012, Experimental Astronomy, 34, 415
Ferrarese, L., & Merritt, D. 2000, Ap. J., 539, L9
Filippenko, A. V., & Ho, L. C. 2003, Ap. J., 588, L13
Frail, D. A., Kulkarni, S. R., Ofek, E. O., Bower, G. C., & Nakar, E. 2012, Ap. J., 747, 70
Gal-Yam, A., Ofek, E. O., Poznanski, D., et al. 2006, Ap. J., 639, 331
Gebhardt, K., Bender, R., Bower, G., et al. 2000, Ap. J., 539, L13
Gezari, S., Heckman, T., Cenko, S. B., et al. 2009, Ap. J., 698, 1367
Gezari, S., Chornock, R., Rest, A., et al. 2012, Nat, 485, 217
Giannios, D., & Metzger, B. D. 2011, Mon. Not. Roy. Astro. Soc., 416, 2102
Granot, J., & Sari, R. 2002, Ap. J., 568, 820
Greene, J. E., & Ho, L. C. 2004, Ap. J., 610, 722
Greiner, J., Schwarz, R., Zharikov, S., & Orio, M. 2000, Astron. Astrophys., 362, L25
Grupe, D., Thomas, H.-C., & Leighly, K. M. 1999, Astron. Astrophys., 350, L31
Guillochon, J., & Ramirez-Ruiz, E. 2013, Ap. J., 767, 25
Hayasaki, K., Stone, N., & Loeb, A. 2013, Mon. Not. Roy. Astro. Soc., 434, 909
Hodge, J. A., Becker, R. H., White, R. L., & Richards, G. T. 2013, Ap. J., 769, 125
Johnston-Hollitt, M., et al., 2015, "Using the SKA Rotation Measure Grid to Reveal the Mysteries of the Magnetised Universe", in "Advancing Astrophysics with the Square Kilometre Array", PoS(AASKA14)092
Kauffmann, G., & Haehnelt, M. 2000, Mon. Not. Roy. Astro. Soc., 311, 576

Kesden, M. 2012, PhRvD, 85, 4037

Khabibullin, I., Sazonov, S., & Sunyaev, R. 2014, Mon. Not. Roy. Astro. Soc., 437, 327

Komossa, S. 2002, Reviews in Modern Astronomy, 15, 27

Komossa, S., & Merritt, D. 2008, ApJ, 683, L21

Komossa, S., et al. 2008, ApJ, 678, L13

Komossa, S. 2012, Advances in Astronomy, 2012, id. 364973

Komossa, S., & Bade, N. 1999, Astron. Astrophys., 343, 775

Krimm, H. A., Kennea, J. A., Holland, S. T., et al. 2011, ATel #3384

Krimm, H. A., Holland, S. T., Corbet, R. H. D., et al. 2013, Ap. J. Suppl., 209, 14

Kroupa, P. 2001, Mon. Not. Roy. Astro. Soc., 322, 231

Levan, A. J, Tanvir, N. R., Cenko, S. B., et al. 2011, Sci, 333, 199

Levan, A. 2012, "Relativistic tidal disruption events", in R. Saxton & S. Komossa (eds.) "Tidal Disruption Events and AGN Outbursts", EPJ Web of Conferences, 39, 2005

Lien, A., Sakamoto, T., Gehrels, N., et al. 2014 Ap. J., 783, 24

Falcke, H., & Biermann, P. L. 1995, Astron. Astrophys., 293, 665

Kumar, P. and Barniol Duran, R. and Bosnjak, Z., et al., 2013, Mon. Not. Roy. Astro. Soc., 434, 3078

Lodato, G., King, A. R., & Pringle, J. E. 2009, Mon. Not. Roy. Astro. Soc., 392, 332

Lodato, G. & Rossi, E. M. 2011, Mon. Not. Roy. Astro. Soc., 410, 359

Loeb, A. 2007, Phys. Rev. D, 99, Issue 4, id. 041103

Maccarone, T. J. 2004, Mon. Not. Roy. Astro. Soc., 351, 1049

Merloni, A., Predehl, P., Becker, W., et al. 2012, ArXiv e-prints, arXiv1209.3114M

Merritt, D., 2009, ApJ, 694, 959

Metzger, B. D., Giannios, D., & Mimica, P. 2012, Mon. Not. Roy. Astro. Soc., 420, 3528

Mooley, Kunal P., Myers, S. T., Hallinan, G., et al. 2014, in AAS Meeting Abstracts, 223

Narayan, R., Igumenshchev, I. V., & Abramowicz, M. A. 2003, Pub. Astron. Soc. Japan, 55 L69

Nandra P., et al., 2013, arXiv1306.2307N

Orosz, G. & Frey, S. 2013, Astron. Astrophys., 553, A13

Paragi, Z., Frey, S., Kaaret, P., et al. 2014, Ap. J., 791, 2

Paragi, Z., Godfrey, L., Reynolds, C., et al. 2014, "Very Long Baseline Interferometry with the SKA", in proceedings of "Advancing Astrophysics with the Square Kilometre Array", PoS(AASKA14)143

Peterson, B. M., Bentz, M. C., Desroches, L.-B., et al. 2005, Ap. J., 632, 799

Phinney, E. S. 1989, Nat, 340, 595

Prandoni, I., et al. 2015, "Revealing the Physics and Evolution of Galaxies and Galaxy Clusters with SKA Continuum Surveys", in proceedings of "Advancing Astrophysics with the Square Kilometre Array", PoS(AASKA14)067

Rees, M. J. 1988, Nat, 333, 523

Reis, R. C., Miller, J. M., Reynolds, M. T., et al. 2012, Sci, 337, 949

Rossi, E. M. & Begelman, M. C. 2009, Mon. Not. Roy. Astro. Soc., 392, 1451

Sadowski, A., Narayan, R., McKinney, J., & Tchekhovskoy, A. 2014, Mon. Not. Roy. Astro. Soc., 439, 503

Sari, R., Kobayashi, S., & Rossi, E. M. 2010, Ap. J., 708, 605

Saxton, C. J., Soria, R., Wu, K., & Kuin, N. P. M. 2012, Mon. Not. Roy. Astro. Soc., 422, 1625
Scott, W. K. 1996, PhD thesis, University Of British Columbia
Shankar, F., Weinberg, D. H., & Miralda-Escudé, J. 2013, Mon. Not. Roy. Astro. Soc., 428, 421
Stone, N., Sari, R., & Loeb, A. 2013, Mon. Not. Roy. Astro. Soc., 435, 1809
Strubbe, L. E., & Quataert, E. 2009, Mon. Not. Roy. Astro. Soc., 400, 2070
Tchekhovskoy, A., Metzger, B. D., Giannios, D., & Kelley, L. Z. 2013, Mon. Not. Roy. Astro. Soc., 437, 2744
Thornton, C. E., Barth, A. J., Ho, L. C., Rutledge, R. E., & Greene, J. E. 2008, Ap. J., 686, 892
van Velzen, S., Farrar, G. R., Gezari, S., et al. 2011, Ap. J., 741, 73
van Velzen, S., Körding, E., & Falcke, H. 2011, Mon. Not. Roy. Astro. Soc., 417, 51
van Velzen, S., & Farrar, G. R. 2014, Ap. J., in press
van Velzen, S., Frail, D., Körding, E., & Falcke, H. 2013, Astron. Astrophys., 552, A5
Volonteri, M. 2010, Astron. Astrophys. Rev., 18, 279
Wang, T.-G., Zhou, H.-Y., Komossa, S., et al. 2012, Ap. J., 749, 115
Zauderer, B. A., Berger, E., Margutti, R., et al. 2013, Ap. J., 767, 152
Zauderer, B. A., Berger, E., Soderberg, A. M., et al. 2011, Nat, 476, 425